\documentclass[epj]{webofc}
\usepackage[varg]{txfonts}   
\usepackage{graphicx}
\usepackage{amssymb}
\usepackage{amsmath}

\woctitle{Wigner 111 -- Colourful \& Deep}
\begin{document}
\title{Universal QGP Hadronization Conditions at  RHIC and LHC}
%
%

\author{%
Johann Rafelski\inst{1}
\and
Michal Petr\' a\v n\inst{1}
}
\institute{Department of Physics, The University of Arizona, Tucson, Arizona 85721, USA}
\abstract{%
We address the principles governing QGP hadronization and particle production in relativistic heavy-ion collisions. We argue that chemical non-equilibrium is required and show that once this condition is assumed a very good description of hadron production in collider  RHIC and at  LHC  heavy ion experiments follows.  We present results of our analysis  as a function of centrality. Comparing most extreme experimental conditions we show that only the reaction volume and degree of strangeness phase space saturation change. We determine the universal QGP fireball hadronization conditions.
}
\maketitle
\section{Hadronization of quark-gluon plasma}
In the hot primordial Universe free color charged quarks and gluons were present in the deconfined phase of matter, the  quark-gluon plasma (QGP). In the laboratory  heavy ion collider RHIC and LHC experiments  we recreate QGP and study this 5th state of  matter. The process QGP undergoes while freezing the color degree of freedom  is called `hadronization': all quarks bind into colorless combinations of three quarks, baryons (and of antiquarks, antibaryons) and quark-antiquark states, mesons. Gluons are either absorbed in the process   balancing color charge, or fuse into quark pairs.

In laboratory collider experiments  the Lorentz contracted heavy ion pancakes encounter each other, partons collide and a droplet of colored particles forms, subsequently thermalizes and expands cooling down, many light quark pairs ($u,d,s$) are produced. The  finite reservoir of energy means that within a relatively short time  the QGP fireball reaches a `color freezing' point where it hadronizes. 

At the time QGP hadronizes quarks and gluons are in kinetic, often also called `thermal', equilibrium. Therefore the directly produced hadrons emerge with an energy  distribution   close to statistical-kinetic equilibrium. However, even when and if QGP is  quark abundance equilibrated, i.e. in chemical equilibrium, in general the produced hadrons cannot be in chemical yield equilibrium. This is so since in general it is impossible that two different phases of matter have similar phase space density for any relevant (nearly) conserved quantity, e.g. baryon number. 

We therefore  evaluate the abundance of all   produced hadrons (including hadron resonances) by considering their phase space distributions including each hadron's  fugacity $\Upsilon_i$ which allows to control their abundance
\begin{equation}
\label{eq:distribution}
 n_i\left( E_i \right) =\frac{g_i}{\Upsilon_i^{-1}\exp\left( E_i/T \right)+S},\quad 
S=\left\{
\begin{array}{ll}
+1 & \; \text{Fermi-Dirac distribution} \\[-1mm]
\phantom{+}0 & \; \text{Boltzmann approximation}\\[-1mm]
-1 & \; \text{Bose-Einstein distribution.} \\[-1mm]
\end{array}
\right.
\end{equation}
Here $g_i$ is the hadron spin degeneracy $g_i=(2J+1)$ and  $E_i = \sqrt{m_i^2+p_i^2}$  hadron energy. $T$ is the temperature of the source, and $S$ distinguishes the appropriate distribution for fermions, bosons and when appropriate the Boltzmann approximation. All different flavor sub-states including isospin states are distinguished and counted separately in our analysis.

Integrating the distribution  Eq.(\ref{eq:distribution}) over the phase space we obtain the total particle multiplicity emerging from the observed volume $V$:
\begin{equation}
\langle N_i \rangle 
=  V \int\frac{\mathrm{d}^3p}{(2\pi)^3} \, n_i
=\frac{g_i V T^3}{2\pi^2}\sum\limits_{n=1}^\infty\frac{(\pm 1)^{n-1}\Upsilon_i^n}{n^3}
\left(\frac{nm_i}{T}\right)^2{K_2}
\left(\frac{nm_i}{T}\right).
\label{eq:yieldexpansion}.
\end{equation}

In principle each hadron has its own chemical non-equilibrium parameter $\Upsilon_i$, Eq.(\ref{eq:distribution}). The statistical hadronization model~\cite{Petran:2013dva} (SHM) approach allows to reduce the large number of parameters: for each hadron species $i$, $\Upsilon_i$  is a product of constituent quark ($u,d,s,c$) fugacities,
\begin{equation}\label{eq:fugacity}
\Upsilon_i = \prod_{f=u,d,s,c} \Upsilon_f,\qquad \Upsilon_f = \gamma_f \lambda_f, \qquad \lambda_f = e^{\mu_f/T}\, \qquad \lambda_{\bar{f}}=\lambda_f^{-1}.
\end{equation}
We see that the fugacity of each constituent quark flavor $\Upsilon_f$ is  composed  of the  product of two factors, and it is important to recognize how the effects of $\lambda$ and $\gamma$ are different: the phase space occupancy $\gamma$ describes the number of quark-antiquark pairs present, whereas $\lambda$ describes the difference in abundance between the quarks and antiquarks of the same flavor.

Because the difference between $u$ and $d$ light quarks is relatively small we assume that the number of $u,\bar u$ and $d,\bar d$ pairs is equal and thus  $\gamma_u=\gamma_d\equiv \gamma_q$. We also introduce  $\lambda_q^2=\lambda_u \lambda_d$ and it is common to introduce the baryochemical potential $\mu_B=3 \ln \lambda_q$. However, we preserve the  net charge per baryon $\langle Q\rangle/\langle B\rangle$ present in the system by fixing a value $\lambda_3=\lambda_u /\lambda_d$. For strange quarks both $\gamma_s\ne 1$ and $\lambda_s\ne 1$ are considered. However we fix the value of  $\lambda_s$ by requiring that   net strangeness $\langle s-\bar s\rangle\to 0$. Nonequilibrium parameters for both strange $\gamma_s$ and light $\gamma_q$ quarks were both first used in analysis of the SPS experimental results~\cite{Letessier:1998sz}. However,  the study of hadron production in QGP hadronization according to direct hadronization process  was first carried out  in Ref.\cite{Koch:1986ud} where furthermore characterization of relative chemical equilibrium and absolute chemical non-equilibrium is found in a format similar to the one introduced here.  

Once we obtain a set of statistical parameters i.e. fugacities $\Upsilon_i$ and temperature $T$ from a best fit to experimental particle abundance data, we can evaluate the hadron yields and the physical properties of the fireball at the instance particle yields are frozen (chemical freeze-out) by summing the contributions of all emerging hadrons $i$ carrying away the energy, entropy, etc.

\section{Hadronization at RHIC and LHC as a function of collision centrality}
We study hadronization process at RHIC and LHC. For both cases considered the data is available in the rapidity interval  $-0.5<y<0.5$.  We use the number of participating nucleons $N_{\rm part}$ from both colliding projectiles to characterize the impact factor of the collision, see Ref.~\cite{Miller:2007ri} for the geometric collision centrality measure overview. At RHIC  we use the STAR experimental yields obtained at $\sqrt{s_{\rm NN}}=62$\, GeV~\cite{Aggarwal:2010ig}:  $\pi^\pm, K^\pm, K_S^0, p^\pm, \phi, \Lambda, \overline{\Lambda},\Xi^\pm$ and $\Omega^\pm$. The LHC Pb--Pb  $\sqrt{s_{\rm NN}}=2760$\,GeV collisions comprise further   K$^*$ and $\phi$ but not  K$_S^0$. The results we discuss here were obtained by the ALICE collaboration and the procedure we used as well as data references are presented in  Ref.\cite{Petran:2013lja}. 

\begin{figure}[t]
\centering
\includegraphics[width=0.75\columnwidth]{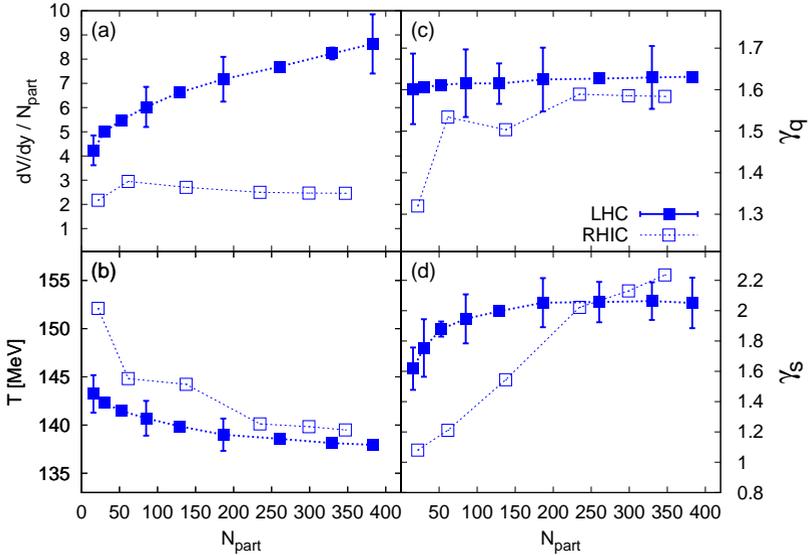}
\caption{\label{fig:VolTemp}SHM parameters obtained at LHC-2760 compared to RHIC-62 as a function $N_{\rm part}$, the collision centrality. Volume normalized to number of participants in panel (a), freeze-out temperature $T$ in panel (b), light quark phase space occupancy $\gamma_q$ in panel (c) and strangeness phase space occupancy $\gamma_s$ in panel (d).}
\end{figure}

Our analysis reveals that the main difference between LHC-2760 and RHIC-62 is about 2 -- 4 times larger volume of the hadronizing fireball, the situation is illustrated in Figure~\ref{fig:VolTemp}a. While the volume $(dV/dy)/N_{\rm part}$ associated with particles observed in the  central rapidity interval $-0.5<y<0.5$ is at RHIC practically constant, at LHC we see a strong nonlinear behavior indicating that new cascading process of entropy formation has been turned on. This is a novel result as prior to LHC experiments all data scaled with $N_{\rm part}$.

The QGP chiral transition temperature obtained in lattice gauge theory is near if not below $T=150$\,MeV~\cite{Borsanyi:2013cga}. The chemical freeze-out  temperature $T$ seen in Figure~\ref{fig:VolTemp}b is found  at LHC for most peripheral collisions $T= 143\pm2$\,MeV, decreasing for most central collision to below $\simeq 140$\,MeV, and systematically somewhat lower compared to RHIC.  This is consistent with  supercooling accompanying the rapid  fireball expansion and dilution, an effect that is expected to increase as the  reaction volume  grows. Therefore these results support sudden hadronization of supercooled  QGP, and we consider the chemical freeze-out to be the same as QGP breakup in RHIC and LHC collisions.

The QGP is characterized by broken color bonds, and thus it is an entropy rich phase requiring in hadronization either: a slow nucleation process which expands the hadronization volume, or the formation  of an overabundance of hadronic  particles, which is expected to be the case for the fast  expanding QGP fireball created in nuclear collision. In Figure~\ref{fig:VolTemp}c we see the light quark phase space occupancy factor $\gamma_q$ which  saturates for all centralities near to the maximum value $\gamma_q\simeq 1.63$, a point near to condensate singularity. This value allows to account for the high entropy content of the color-bond broken QGP phase. The global chemical non-equilibrium is in agreement with the direct hadronization of QGP into free-streaming particles at RHIC and LHC as implied by above consideration of supercooling.

Strangeness abundance has been predicted to be the characteristic signature of QGP formation~\cite{Rafelski:1982ii}. The  strangeness phase space factor $\gamma_s$  seen  in Figure~\ref{fig:VolTemp}d  indicates a large overpopulation at both RHIC and LHC. At LHC the rise occurs  already for relatively peripheral collisions with $N_{\rm part}<150$ as a function of centrality and the value saturates at $\gamma_s<2.1$. The early saturation at LHC can be understood as being due to more extreme initial temperature, which  speeds up strangeness production. At RHIC   there is no evidence of strangeness saturation. The significant difference in strangeness evolution between LHC and RHIC is like the hadronization volume  controlling final particle abundances.

\begin{figure}[t]
\centering
\includegraphics[width=0.7\textwidth]{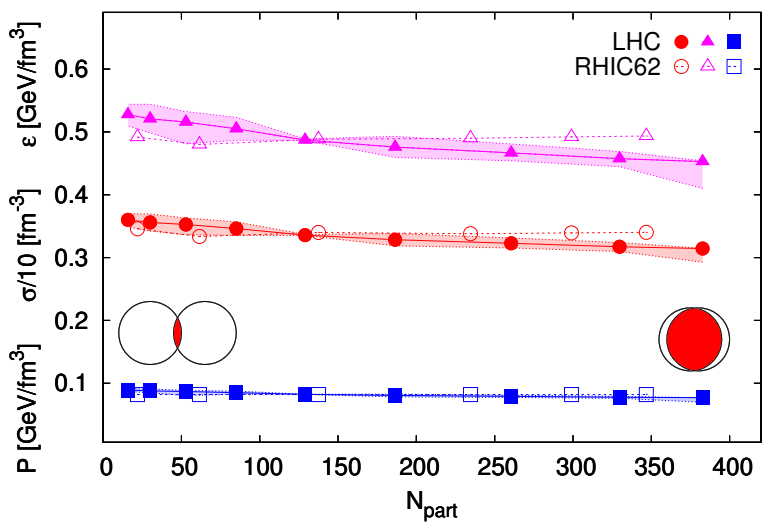}
\caption{\label{fig:BulkProperties}Bulk properties of the hadronizing  fireball obtained at LHC-2760 and compared to RHIC-62 as a function $N_{\rm part}$. From top to bottom: energy density $\varepsilon$, entropy density $\sigma$ (scaled down by factor 10),   pressure $P$.}
\end{figure}

Our analysis demonstrates the concordance of LHC and RHIC physical QGP  properties, which we show in Figure~\ref{fig:BulkProperties}. The three results correspond from  top to bottom to: energy density $\varepsilon=0.50\pm 0.05\,\mathrm{GeV}$, entropy density $\sigma = 3.35 \pm 0.30\,\mathrm{fm}^{-3}$, and  pressure $P=82\pm 8\,\mathrm{MeV/fm}^3$. Identification of the same fireball properties at LHC as at RHIC, i.e.  presence of  universal hadronization conditions of the QGP fireball~\cite{Petran:2013qla}, proposed already comparing  SPS with RHIC experiments~\cite{Rafelski:2009jr}  means that for vastly differing initial conditions the physical principles governing creation of final state hadrons are universal, a situation most easily associated with the direct breakup of supercooled 5th phase of matter, QGP, into final state hadrons.

\begin{thebibliography}{9}

\bibitem{Petran:2013dva} 
  M.~Petran, J.~Letessier, J.~Rafelski and G.~Torrieri,
  Comput.\ Phys.\ Commun.\  {\bf 185}, 2056 (2014). 

\bibitem{Letessier:1998sz} 
  J.~Letessier and J.~Rafelski,
  Phys.\ Rev.\ C {\bf 59}, 947 (1999),
and 
  J.\ Phys.\ G {\bf 25}, 295 (1999).

\bibitem{Koch:1986ud} 
  P.~Koch, B.~Muller and J.~Rafelski,
  Phys.\ Rept.\  {\bf 142}, 167 (1986).

\bibitem{Miller:2007ri} 
  M.~L.~Miller, K.~Reygers, S.~J.~Sanders and P.~Steinberg,
  Ann.\ Rev.\ Nucl.\ Part.\ Sci.\  {\bf 57}, 205 (2007).

\bibitem{Aggarwal:2010ig} 
  M.~M.~Aggarwal {\it et al.}  [STAR Collaboration],
  Phys.\ Rev.\ C {\bf 83}, 024901 (2011).

\bibitem{Petran:2013lja} 
  M.~Petrán, J.~Letessier, V.~Petráček and J.~Rafelski,
  Phys.\ Rev.\ C {\bf 88},  034907 (2013).

\bibitem{Borsanyi:2013cga} 
  S.~Borsanyi, Z.~Fodor, C.~Hoelbling, S.~D.~Katz, S.~Krieg and K.~K.~Szabo,
  PoS {\small  LATTICE} {\bf 2013}, 155 (2013),
and 
  Phys.\ Lett.\ B {\bf 730}, 99 (2014).

\bibitem{Rafelski:1982ii} 
  J.~Rafelski,
  Phys.\ Rept.\  {\bf 88}, 331 (1982).


\bibitem{Petran:2013qla} 
  M.~Petran and J.~Rafelski,
  Phys.\ Rev.\ C {\bf 88}, no. 2, 021901 (2013).

\bibitem{Rafelski:2009jr} 
  J.\,Rafelski and J.\,Letessier,
  J.\,Phys.\,G {\bf 36}, 064017 (2009), and 
  PoS {\small CONFINEMENT} {\bf 8}, 111\,(2008).

\end{thebibliography}
\end{document}